\documentclass{iau-JDSS}
\usepackage{graphicx}

\title[A Large X-ray Sample of Fossil Groups]
{A Large X-ray Sample of Fossil Groups}

\author[Miller et al.]   
{Eric D. Miller$^1$, Eli Rykoff$^2$, Renato de Alencar Dupke$^{3,4}$, 
Claudia Mendes de Oliveira$^5$, Timothy McKay$^4$, 
Benjamin Koester$^6$}

\affiliation{$^1$MIT Kavli Institute,
$^2$UCSB,
$^3$ON/MCT,
$^4$U.~Michigan,
$^5$USP,
$^6$U.~Chicago}

\pubyear{2009}
\volume{Volume 15}  
\pagerange{119--126}
\date{?? and in revised form ??}
\jname{Highlights of Astronomy, Volume 15}
\editors{Ian F Corbett, ed.}

\begin{document}

\maketitle

\begin{abstract}
We present Chandra snapshot observations of the first large X-ray sample
of optically identified fossil groups.  For 9 of 14 candidate groups, we
are able to determine the X-ray luminosity and temperature, which span a
range typical of large ellipticals to rich groups of galaxies.  We discuss
these initial results in the context of group IGM and central galaxy ISM
evolution, and we also describe plans for a deep X-ray follow-up program.
\end{abstract}


Fossil groups (FGs) are systems dominated by a single, giant elliptical
galaxy, yet their X-ray emission indicates a deeper cluster-scale
gravitational potential.  They are thought to be old, undisturbed galaxy
groups, however these systems may be younger or more active than previously
thought (see Dupke et al.~in these proceedings).  These results are
complicated by the small number of FGs with deep X-ray data. 

\begin{figure}[b]
\begin{centering}
\begin{minipage}[l]{.48\linewidth}
\includegraphics[height=\linewidth,angle=270]{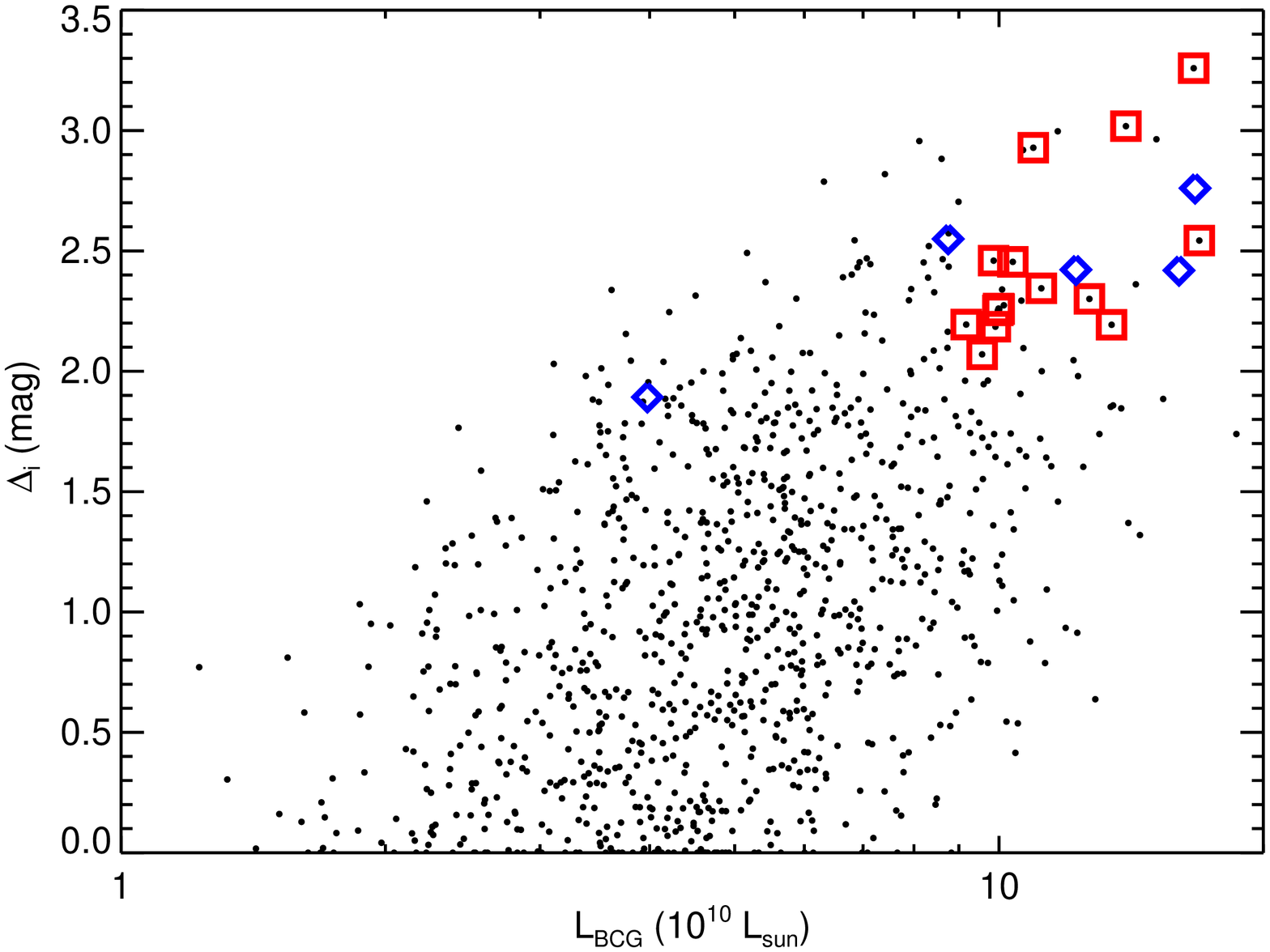}
\caption{Magnitude differential vs.~BCG luminosity for all maxBCG
clusters with $9 < N_{200} < 25$; open squares identify the 15 FG
candidates.  Diamonds show known FGs.}
\label{fig1}
\end{minipage}
\begin{minipage}[c]{.04\linewidth}
\end{minipage}
\begin{minipage}[r]{.48\linewidth}
\includegraphics[width=\linewidth]{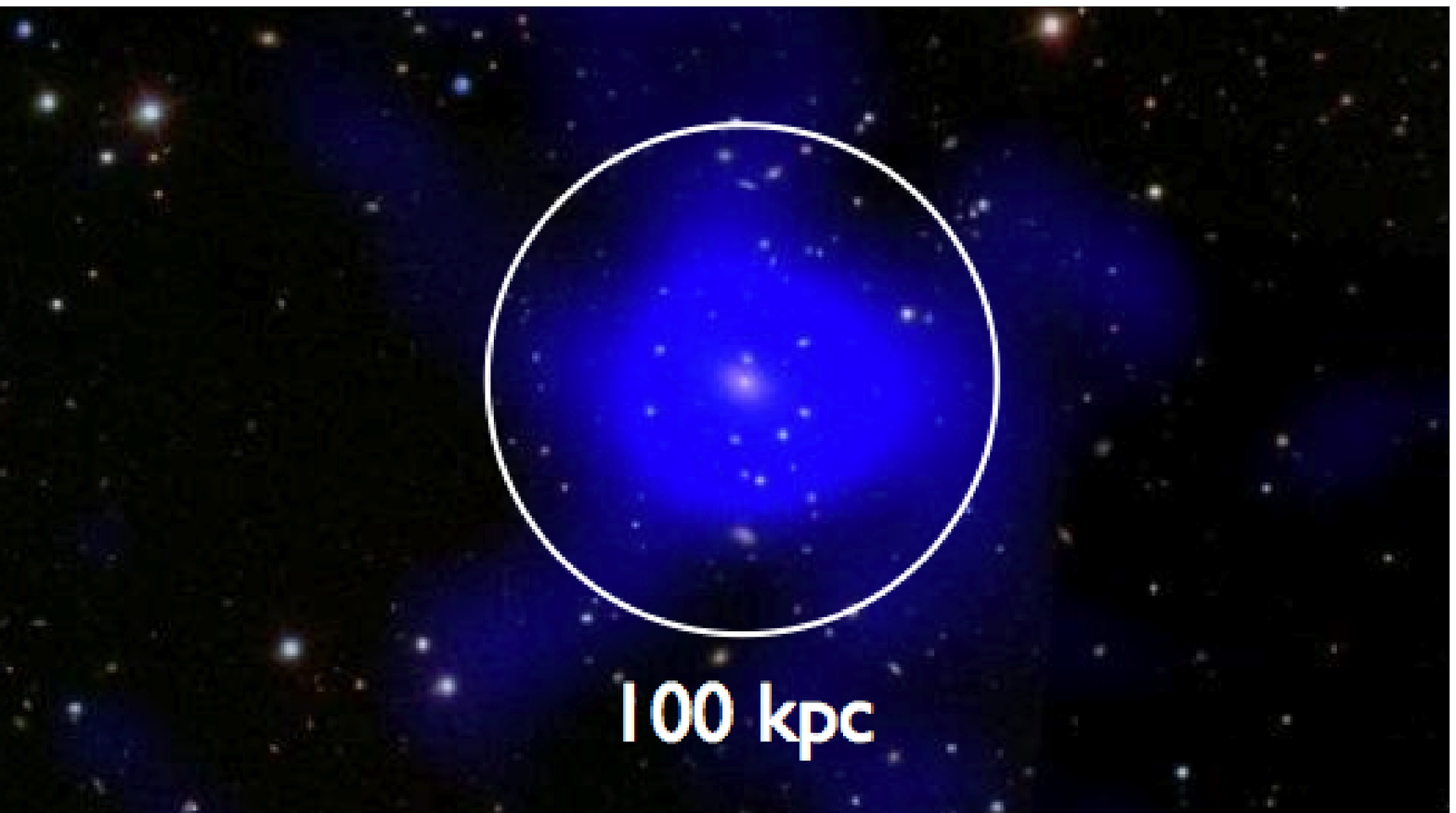}
\vspace*{\baselineskip}
\caption{SDSS J0856+0553, a $z = 0.09$ FG candidate.  The X-ray image is
plotted over the SDSS $g,r,i$ composite image.}
\label{fig2}
\end{minipage}
\end{centering}
\end{figure}

To address this, we have constructed a sample of 15 FG
candidates from the maxBCG cluster catalog (Koester et al.~2007), using
the criteria $0.09 < z < 0.15$, $L_{\rm BCG} > 9\times10^{11} L_{\odot}$,
and $\Delta i > 2.0$ between the BCG and second ranked galaxy within
$R_{200}/2$ (see Figure \ref{fig1}).  We have obtained 5--10 ksec
\textit{Chandra} snapshot observations of 14 targets, and we detect diffuse
X-ray emission from 11 of them at $> 90\%$ confidence, measuring $T_X$ for 9
of these.  One detection is shown in Figure \ref{fig2}.  
The measured $L_X$ and $T_X$ are similar to what is expected for groups of
galaxies.  Deep follow-up with \textit{XMM} is necessary to measure $T_X$
profiles, surface brightness profiles, concentration, and abundances,
thereby constraining the formation mechanism of these peculiar but numerous
systems.

\vspace*{-\baselineskip}

\end{document}